\documentclass[prl,showpacs,twocolumn, preprintnumbers,amsmath,amssymb]{revtex4}
\usepackage{amsmath}
\usepackage{amssymb}
\usepackage{graphicx}
\usepackage{dcolumn}

\newcommand{\Kn}{\ensuremath{{\rm Kn}}}

\newcommand{\Ma}{\ensuremath{{\rm Ma}}}

\newcommand{\bc}{\ensuremath{\mathbf{v}}}

\newcommand{\cs}{c_{\rm s}}

\newcommand{\jj}{{\mbox{\boldmath$j$}}}

\newcommand{\xx}{\mbox{\boldmath$x$}}

\newcommand{\vel}{\mbox{\boldmath$c$}}
\newcommand{\uu}{\mbox{\boldmath$u$}}

\begin{document}
\title
{Consistent Lattice Boltzmann Method}
\author{Santosh Ansumali and Iliya V. Karlin}
\affiliation {ETH-Z\"urich,  Institute of Energy Technology,
CH-8092 Z\"urich, Switzerland}

\begin{abstract}
The problem of energy conservation in the lattice Boltzmann method
is solved. A novel model with energy conservation is derived from
Boltzmann's kinetic theory. It is demonstrated that the full
thermo-hydrodynamics pertinent to the Boltzmann equation is
recovered in the domain where variations around the reference
temperature are small. Simulation of a Poiseuille micro-flow is
performed in a quantitative agreement with exact results for low
and moderate Knudsen numbers. The new model extends in a natural
way the standard lattice Boltzmann method to a thermodynamically
consistent simulation tool for nearly-incompressible flows.
\end{abstract}
\pacs{05.20.Dd, 47.11.+j}
\date{\today}
\maketitle

The overwhelming majority of fluid flows of physical and
engineering interest are slow. That is, characteristic flow speed
$u$ is small compared to the speed of sound $\cs$. This is
quantified by the Mach number, $\Ma\sim u/\cs$. In typical
situations, $\Ma$ varies from $10^{-3}-10^{-2}$ in hydrodynamic
flows (turbines, reactors etc) to $10^{-4}$ in flows at a
micrometer scale. In this paper we address a wide class of flows
at $\Ma\ll 1$. Then the simplest characterization of the degree of
molecularity  is the Knudsen number $\Kn\sim \lambda/H$, the ratio
of the mean free path of molecules $\lambda$ and the
characteristic scale $H$ of variation of hydrodynamic fields
(density, momentum, and energy). When $\Kn\lesssim 10^{-3}$, one
considers the hydrodynamic limit where molecularity reduces to
specific for each molecular model set of transport coefficients
(viscosity, thermal conductivity etc). If, in addition, the Mach
number is also small, one enjoys the incompressible hydrodynamics
with the ordering $\Kn\ll\Ma\ll1$, and the flow can be
characterized solely by the ratio ${\rm Re}\sim\Ma/\Kn$ (one of
the definitions of the Reynolds number).

Contemporary computational fluid dynamics becomes increasingly
more interested in the domain where Mach number remains small but
Knudsen number increases, thus, the incompressibility  becomes
gradually lost. Because of its relevance to the engineering of
micro electro-mechanical systems (MEMS), the branch of
computational fluid dynamics focused on micro scale phenomena is
often called ``micro-fluidics" \cite{Karniadakis2}. Typical flows
in micro-devices are highly subsonic (with characteristic flow
velocities about $0.2\ m/s$, corresponding to ${\rm Ma} \sim
10^{-4}$), while Knudsen number varies from ${\rm Kn}\sim 10^{-2}$
(so-called slip-flow regime) to ${\rm Kn} \sim 1$ (moderately
rarefied gas flows) \cite{Karniadakis2}. There is much need for
computational models in the domain of slow flows where the effects
of molecularity become increasingly more pronounced.

In recent years, the lattice Boltzmann method has drawn
considerable attention as a simulation method for flows at low
Mach numbers. Especially popular are the so-called isothermal
lattice Boltzmann models  (ILBM) without energy conservation
\cite{succi}. Recently, there was increasing interest in applying
these models also for micro-flow simulations
\cite{Karniadakis2,AK4,NIE,SLIP,KWOK,ELBMMICRO,AKFB05}. The
hydrodynamic (locally conserved) fields in the ILBM are the
density $\rho$ and the momentum density $\jj$, whereas the
conservation of the energy is not addressed. Construction of the
ILBM may vary among the authors, but all these models have one
point in common: The lack of energy conservation inevitably leads
to a bulk viscosity. Indeed, the non-equilibrium part of the
stress tensor in ILBM reads:

\begin{equation}
P_{\alpha\beta}^{\rm neq} \sim {\rm Kn}\left[ \partial_{\alpha}
\left(\frac{j_\beta}{\rho}\right) +
    \partial_{\beta} \left(\frac{j_{\alpha}}{\rho}\right)
    \right].\label{StressLB}
\end{equation}
This tensor is not traceless, $P_{\alpha\alpha}^{\rm neq} \sim 2
{\rm Kn} \partial_{\alpha}(j_\alpha/\rho)$, which immediately
leads to the bulk viscosity terms in the equation for the momentum
density.  We remind that the physical bulk viscosity in
hydrodynamic models is related to a redistribution of the energy
among the translational and internal degrees of freedom of
molecules rather than to any non-conservation of the energy. Thus,
from the physical standpoint, the bulk viscosity of the ILB models
is spurious. Certainly, the presence of the bulk viscosity,
spurious or not, by no means precludes the limit of incompressible
hydrodynamics  because, loosely speaking, the divergence of the
velocity field $\uu=\jj/\rho$ vanishes in that limit
\cite{AKOe05}. Thus, ILBM is a valid model for the incompressible
hydrodynamics. However, the spurious bulk viscosity of ILBM
becomes a severe drawback when such models are applied to
weakly-compressible or micro-flow simulations.

The best way to illustrate this problem is to consider a
representative example. Plane Poiseuille flow is one of the most
studied benchmarks on gas dynamics. The gas moves between two
parallel plates driven by a fixed pressure difference between the
inlet and outlet. It is known from the classical kinetic theory
\cite{Cerci} that the flow rate $Q$ has the following asymptotic
at low and high Knudsen numbers:

\begin{eqnarray*}
  Q_0 &=& (6 {\rm Kn})^{-1}+s+(2s^2-1)\;{\rm Kn},\ \Kn\ll1 \\
  Q_{\infty} &\sim& (1/\sqrt{\pi})\ln{({\rm Kn})} + O(1),\ \Kn \to
  \infty,
\end{eqnarray*}
with $s=1.015$.
 These two asymptotic limits ensure that the flow
rate has a minimum at some finite $\rm Kn$ (the Knudsen minimum
\cite{Knudsen}).
While a qualitative agreement of the ILBM simulations with the
continuous-velocity kinetic theory was found at all Knudsen
numbers \cite{Toschi05,AKFB05}, the quantitative agreement is poor
beyond the slip-flow regime at ${\rm Kn}>10^{-2}$. It was found
that the ILBM {\it systematically over-predicts} the flow rate at
{\it small} Knudsen numbers. This is the effect of the bulk
viscosity which can be qualitatively explained as follows: At low
Knudsen numbers the behavior is still dominated by particle's
collisions in the bulk, therefore, the steady state is reached
upon a balance between the frictional force $\sim {\rm
Kn}\partial_{\beta}P^{\rm neq}_{\alpha\beta}$ and the forcing due
to the constant pressure gap between the inlet and the outlet.
Therefore, if there is additional contribution of the bulk
viscosity (more friction), this balance at the same ${\rm Kn}$
shifts to a higher velocity at the steady state, and will result
in the over-prediction of the flow rate. On the other hand, the
lack of the energy conservation also contributes to the
over-prediction at high Knudsen numbers by a different mechanism:
Since collisions in the bulk become rear, it becomes important
that the energy be correctly redistributed in these rear events.

In this paper we introduce  new lattice Boltzmann models with the
energy conservation. These models are derived from the continuous
kinetic theory, are free from the drawbacks of the isothermal
lattice Boltzmann models, and at the same time they retain in full
the outstanding computational efficiency of the latter.

The structure of the paper is as follows: First, we shall derive
the discrete-velocity model from the continuous-velocity kinetic
theory. Second, we shall redo the computation of the
micro-Poiseuille flow with the new model, and demonstrate a
significant (roughly, order of magnitude in ${\rm Kn}$)
improvement in the accuracy with respect to isothermal simulation.
Third, we shall explain in which points the present derivation
differs from the earlier studies. A brief discussion concludes the
paper.

Starting point of our derivation is the grand canonic potential of
the Boltzmann kinetic theory,

\begin{equation}
H=\int F\ln{F}d\bc+\mu\int F d\bc+ \zeta_{\alpha}\int Fv_{\alpha}
d\bc + \gamma\int F v^2 d\bc, \label{ContH}
\end{equation}
where $F(\xx,\bc)$ is the one-particle distribution function, and
$\mu$, $\zeta_{\alpha}$, and $\gamma$ are Lagrange multipliers
corresponding to density, momentum and energy, respectively. The
$D+2$-parametric family of functionals (\ref{ContH}), where $D$ is
the dimension of the velocity space, describes the equilibrium
states as its minima, $\delta H=0$, and it also defines the
locally conserved fields (density $\rho$, momentum $\jj$, and
energy $e$),
\begin{equation}
\frac{\partial H}{\partial \mu}=\rho,\ \frac{\partial H}{\partial
\zeta_{\alpha}}=j_{\alpha},\ \frac{\partial H}{\partial \gamma}=e.
\label{DefinitionHydro}
\end{equation}

In order to derive the discrete velocity kinetic theory,  the
functional (\ref{ContH}) is evaluated with the help of the
$D$-dimensional Gauss-Hermite quadrature with the Gaussian weight
$W=\left(2\pi
\theta_0\right)^{-D/2}\exp{\left(-(mv^2)/(2\theta_0)\right)}$,
where $\theta_0=(k_{\rm B}T_0/m)$ is the reduced uniform reference
temperature. We remind that the quadrature evaluation of an
integral replaces it by a sum, $\int
W(\bc)G(\bc)d\bc\approx\sum_{i=1}^{n_{\rm d}}W_iG(\bc_i)$, where
$\bc_i$, $i=1,\dots,n_{\rm d}$ are the nodes (or abscissas) of the
quadrature, and $W_i$ are corresponding weights. In the case under
consideration, the nodes of the quadrature (discrete velocities)
are situated at the zeroes of Hermite polynomials.

We now proceed with the Gauss-Hermite quadrature evaluation of the
velocity integral (\ref{ContH}). For concreteness, we shall
consider the third-order Hermite polynomial. Then $n_{\rm d}=3^D$,
and the discrete velocities and weights is constructed as follows:
For $D=1$, the three roots and corresponding weights are
$(-\sqrt{3\theta_0},0,\sqrt{3\theta_0})$, $(1/6,2/3,1/6)$; for
$D>1$, the roots are all possible tensor products of the roots in
$D=1$, and the weights are corresponding products of
one-dimensional weights. We shall consider $D=3$ below, that is
$n_{\rm d}=27$ (same considerations apply to any quadrature, in
particular, to the popular $9$-velocity model for $D=2$). As is
well known, the third-order quadrature has the unique feature that
its nodes form a face-centered square lattice which is the crucial
feature to the further lattice Boltzmann discretization in space
and time. Introducing the populations,
 $f_i=W_i(2\pi
\theta_0)^{3/2}\exp{\left(v_i^2/(2\theta_0)\right)}F(\xx,\bc_i)$,
and using the reduced discrete velocities,
$\vel_i=\bc_i/\sqrt{3\theta_0}$, we write the quadrature for
(\ref{ContH})
\begin{equation}
\label{H} H=\sum_{i=1}^{27}\left\{ f_{i}\ln
\left(\frac{f_{i}}{W_i}\right) +\mu f_i +\zeta_{\alpha}f_ic_i+
\gamma f_i c_i^2\right\}.
\end{equation}
Differentiation of (\ref{H}) with respect to Lagrange multipliers
defines the locally conserved fields in the discrete case (cf.\
(\ref{DefinitionHydro})),
\begin{equation}\label{Fields27}
\sum_{i=1}^{27} \{ 1,\, c_{i\alpha},\, c_i^2 \} f_i =\{\rho,\,
j_{\alpha}, \, 3p+ \rho^{-1}j^2 \}.
\end{equation}
The equilibria $f_i^{\rm eq}$ are now found as minima of $H$
(\ref{H}). From the extremum condition, $\delta H=0$, it follows

\begin{equation}
f_i^{\rm eq}=W_i\exp\{-\mu-\zeta_{\alpha}c_{i\alpha}-\gamma
c_i^2\}. \label{EquilibriumImplicit}
\end{equation}
In order to express the Lagrange multipliers in
(\ref{EquilibriumImplicit}) in terms of hydrodynamic fields
(\ref{Fields27}), we substitute (\ref{EquilibriumImplicit}) into
(\ref{Fields27}) and derive the functions $\mu(\rho, \jj, p)$,
$\zeta_{\alpha}(\rho, \jj, p)$ and $\gamma(\rho, \jj, p)$ by
perturbation for small momentum, owing for the fact that
$\zeta_{\alpha}(\rho, \mathbf{0}, p)=0$, and that the
zero-momentum functions $\mu(\rho, \mathbf{0}, p)$ and
$\gamma(\rho, \mathbf{0}, p)$ can be found in a closed form.
Computation is quite straightforward, and we write here the final
result to second order in the momentum:

\begin{widetext}
\begin{equation}
f_i^{\rm eq}(\rho,\jj,p)=\rho\left(1-\frac{p}{\rho}\right)^3
     \left(\frac{{\frac{p}{\rho}}}{2\left(1 - {\frac{p}{\rho}}\right)}\right)^{c_i^2}
     \left[1+ \frac{c_{i\alpha }j_{\alpha}}{p}
      +\frac{j_{\alpha}j_{\beta}}{2p^2}
     \left(c_{i\alpha }c_{i\beta }
     -
     \frac{6\frac{p^2}{\rho^2}+c_i^2\left(1 - 3\,  \frac{p}{\rho} \right)}{
3\left( 1 - \frac{p}{\rho} \right)}\delta _{\alpha\beta}
            \right)
    \right].
    \label{Equilibrium27vThermal}
\end{equation}
\end{widetext}
The pre-factor in this formula has the following limit when
$(p/\rho)\to (1/3)$:
\begin{equation}
\lim_{(p/\rho)\to (1/3)}\left(1-\frac{p}{\rho}\right)^3
     \left(\frac{{\frac{p}{\rho}}}{2\left(1 -
     {\frac{p}{\rho}}\right)}\right)^{c_i^2}=W_i.
     \label{PrefactorLimit27}
\end{equation}
The implication of this limit will be important below when we will
be discussing the relation of the present model to the isothermal
lattice Boltzmann model.

We now proceed with the evaluation of the stress tensor $P^{\rm
eq}_{\alpha\beta}(\rho,\jj,p)$ and of the energy flux $q^{\rm
eq}_{\alpha}(\rho,\jj,p)$ at equilibrium. The important
observation to be made here is that if the pressure to density
ratio satisfies the condition, then $P^{\rm eq}_{\alpha\beta}$ and
$q^{\rm eq}_{\alpha}$ satisfy the corresponding relations
pertinent to the continuous-velocity Maxwell distribution. In
dimensional units, the condition just mentioned reads $p=(k_{\rm
B}T_0\rho)/m$, that is, it corresponds to the ideal gas equation
of state at the reference temperature of the Gaussian weight.
Moreover, if we allow small variations of the pressure around the
point $p/\rho=1/3$, namely, $|(p/\rho)-(1/3)|\sim {\rm Ma}^2$, the
Maxwell's form of the functions $P^{\rm eq}_{\alpha\beta}$ and
$q^{\rm eq}_{\alpha}$ persists, and we have

\begin{eqnarray}
P_{\alpha\beta}^{\rm eq} &= & \sum_{i=1}^{27}f^{\rm
eq}_ic_{i\alpha}c_{i\beta}=p \delta_{\alpha \beta} +
\frac{j_{\alpha} j_{\beta}}{\rho}, \\
q^{\rm eq}_{\alpha} &= &\sum_{i=1}^{27}f^{\rm
eq}_ic_{i\alpha}c_{i}^2=5\frac{p}{\rho}j_{\alpha}.
\end{eqnarray}
Note that the variation of the pressure to density ratio of the
order $\Ma^2$ is pertinent to flow phenomena where the dominant
effect on the temperature is the viscous dissipation and
associated heat conduction. Condition $|(p/\rho)-(1/3)|\sim {\rm
Ma}^2$ is a conservative estimate of the domain of validity of the
present model.

With the equilibrium (\ref{Equilibrium27vThermal}), we write up
the simplest kinetic equation (the Bhatnagar-Gross-Krook model),
\begin{equation}
\label{LBMcont27} \partial_t f_i+c_{i\alpha}\partial_{\alpha}f_i=
-\frac{1}{\tau}(f_i-f_i^{\rm eq}(\rho, \jj, p)),
\end{equation}
where $\tau>0$ is the relaxation time. In order to find out the
hydrodynamic limit of the model, we perform the Chapman-Enskog
analysis at low Mach numbers. In so doing, we neglect all terms in
$j_{\alpha}$ of the order three and higher, and we end up with the
following non-equilibrium expressions for the stress and the heat
flux:

\begin{eqnarray}
P_{\alpha\beta}^{\rm neq} = - \tau p\left[ \partial_{\alpha}
\left(\frac{j_\beta}{\rho}\right) +
    \partial_{\beta} \left(\frac{j_{\alpha}}{\rho}\right)-\frac{2}{3}\delta_{\alpha\beta}
    \partial_{\gamma} \left(\frac{j_{\gamma}}{\rho}\right)
    \right],\label{Stress27ThermalNoneq}\\
q_{\alpha}^{\rm neq} = -2\tau p
 \partial_{\alpha}\left(\frac{p}{\rho}\right).\label{Heat27ThermalNoneq}
\end{eqnarray}
The most important achievement is that the non-equilibrium
(Newtonian) stress (\ref{Stress27ThermalNoneq}) is traceless, that
is, by preserving the energy conservation we eliminated the
spurious bulk viscosity. The heat flux (\ref{Heat27ThermalNoneq})
obeys the Fourier law.

\begin{center}
\begin{figure}[t]
\includegraphics[scale=0.30]{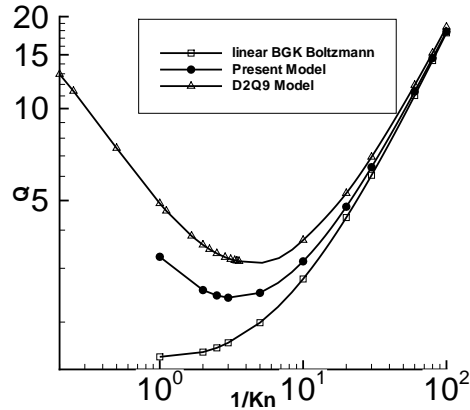}
\caption{\label{Fig1} Flow rate in the pressure driven Poiseuille
flow as a function of inverse Knudsen number. Comparison of the
present energy-conserving model with the isothermal lattice
Boltzmann model $D2Q9$ \cite{AK5} and the continuous linearized
Boltzmann-BGK model \cite{CerciVar}.} \label{Fig1}
\end{figure}
\end{center}

We have implemented the lattice Boltzmann space-time
discretization of the kinetic equation (\ref{LBMcont27}), and redo
the micro-Poiseuille flow simulation mentioned in the
introduction. Results are presented in Fig.\ \ref{Fig2}, where the
present model is compared to the exact solution of the continuous
linearized BGK model \cite{CerciVar}, and the $2DQ9$ isothermal
lattice Boltzmann model with the spurious bulk viscosity
\cite{AK5}. It is clearly visible that the effect of the spurious
bulk viscosity of the isothermal model is completely eliminated,
and that the quantitative agreement with the continuous BGK model
extends up to ${\rm Kn}=0.1$.

Finally, let us place our derivation with respect to previously
reported lattice Boltzmann models on the same lattice. If we
substitute $p=(1/3)\rho$ into the equilibrium function
(\ref{Equilibrium27vThermal}), and use the limit
(\ref{PrefactorLimit27}), then $f^{\rm eq}_i(\rho,\jj,\rho/3)$
recovers the second-order polynomial equilibrium of the isothermal
lattice Boltzmann method on the same lattice \cite{LBGK1,LBGK2},
and instead of the traceless stress tensor
(\ref{Stress27ThermalNoneq}) we recover (\ref{StressLB}) with the
bulk viscosity component. It needs to be stressed that the
second-order polynomial in $\jj$ (\ref{Equilibrium27vThermal}) is
an approximation to the positive-definite discrete-velocity
equilibrium (\ref{EquilibriumImplicit}). Same as with the
isothermal models, this second-order approximation simply happens
to be good enough for stable computations at ${\rm Ma}<0.1$. If we
keep the relation $p=(1/3)\rho$ to all the orders in $\jj$, we
recover the exact positive-definite equilibrium of the isothermal
$27$-velocities entropic lattice Boltzmann model \cite{AK5}:

\begin{widetext}
\begin{equation}
\label{TED}
 f^{\rm eq}_i(\rho,\jj,\rho/3)=\rho W_i\prod_{\alpha=1}^{3}
\left(2 -\sqrt{1+ 3 {\left(j_{\alpha}/\rho\right)^2}}
\right)\,\left(\frac{2\,\left(j_{\alpha}/\rho\right) + \sqrt{1+
3\,\left(j_{\alpha}/\rho\right)^2}}{1-\left(j_{\alpha}/\rho\right)}
\right)^{c_{i\alpha}}.
\end{equation}
\end{widetext}
Entropic stabilization procedure \cite{ELB1,DHT,Boghosian} can be
applied for the present model in order to achieve small values of
the transport coefficients but we do not address this here.

Our approach to the discretization of the velocity space differs
from the earlier considerations \cite{HeLuo,ShanHe}. While we use
the same Gauss-Hermite quadrature, we apply it on the grand
canonical potential (\ref{ContH}) (that is, we evaluate the
velocity integral (\ref{ContH}) as pertinent to the meaning of a
quadrature), and after that find the discrete-velocity equilibrium
upon minimization of the discrete-velocity grand canonical
potential (\ref{H}). Instead, authors of \cite{HeLuo,ShanHe}
evaluate the local Maxwellian (that is, they evaluate a  {\it
function}, not an integral) of continuous kinetic theory,
$M(\rho,\jj,T;\bc)=\rho(2\pi k_{\rm
B}T/m)^{-D/2}\exp(-m(\bc-\uu)^2/2k_{\rm B}T)$, $\uu=\jj/\rho$, on
the nodes of the quadrature. Certainly, just replacing
$M(\rho,\jj,T;\bc)\to M_i(\rho,\jj,T;\bc_i)$ makes no sense
because the conservation laws will be lost. However, it was
noticed in \cite{HeLuo} that when the second-order expansion of
the Maxwellian is used instead, $M^{(2)}=\rho
W(1+av_{\alpha}j_{\alpha}+bv_{\alpha}v_{\beta}j_{\alpha}j_{\beta})$,
the replacement, $M^{(2)}\to\rho
W_i(1+av_{i\alpha}j_{\alpha}+bv_{i\alpha}v_{i\beta}j_{\alpha}j_{\beta})$,
coincides with the previously known second-order equilibrium of
the isothermal lattice Boltzmann method \cite{LBGK1,LBGK2}. It
should be stressed that while in \cite{HeLuo,ShanHe} the
Maxwellian {\it must} be truncated to second order (in order to
rescue conservation laws), and thus the positivity of populations
has to be sacrificed together with the second law of
thermodynamics (Boltzmann's $H$-theorem), our equation
(\ref{Equilibrium27vThermal}) is just a good approximation to the
positive equilibrium (\ref{EquilibriumImplicit}), and, if
required, further terms can be computed in order to maintain
positivity of the equilibrium populations to any required degree
of accuracy. The discretization of the velocity space done at the
level of generating functional (\ref{ContH}) obviously violates
none of the properties of the continuous kinetic theory.

Ref.\ \cite{ShanHe} indicated that the nodes of the fourth-order
quadrature can be used for establishing a thermal model. Such
model was indeed constructed and implemented in \cite{AK5,Adiss}.
However, even though the admissible temperature variation in this
model is larger,  it is bound to be less efficient than the
lattice Boltzmann method. In that sense, the lattice Boltzmann
model with energy conservation derived in this paper is a good
compromise for simulations of almost-isothermal low Mach number
flows.

In conclusion, we have given a microscopic derivation of a genuine
lattice Boltzmann model with energy conservation for simulations
of incompressible and almost-incompressible flows. This model is a
natural extension of the previously known isothermal models on the
same lattices, and, due to the reasons explained above, it was
entirely ``overlooked" in all previous derivations. While we have
considered here the $27$-velocities lattice only, we shall address
other important cases such as the $15$- and $19$-velocities $3D$
lattices  and the $9$-velocity $2D$ lattice in our subsequent
publications. The new models are as efficient as the previous
isothermal lattice Boltzmann models on the same lattices, and at
the same time they extend considerably the domain of validity of
lattice Boltzmann computations especially into the micro-flow
domain. As we have already said, in principle, the spurious bulk
viscosity of ILBM is not an obstacle for using them for
incompressible hydrodynamics simulation. However, even in that
case, the present models can be preferred on the ground that they
correspond more to the physics. The approach to the discretization
of the microscopic theories uses explicitly the Legendre structure
of thermodynamics \cite{GKbook}, and can be used in other problems
of reducing description. This work was supported by the
BFE-Project Nr. 100862.

\bibliography{mmd}

\end{document}